\documentclass{iau}
\usepackage{graphicx}
\usepackage{amsmath}
\usepackage{amssymb}

\title[Automatic stellar parameterization pipeline for LAMOST] 
{Automatic stellar spectral parameterization pipeline for LAMOST survey}

\author[Y. Wu et al.]   
{Yue Wu, Ali Luo, Bing Du, Yongheng Zhao and Hailong Yuan
 }

\affiliation{Key Laboratory of Optical Astronomy, National Astronomical Observatories, Chinese
Academy of Sciences, Beijing 100012, China \\email: {\tt wuyue@bao.ac.cn}
}

\pubyear{2014}
\volume{306}  
\pagerange{119--126}
\jname{Statistical Challenges in 21st Century Cosmology}
\editors{Alan Heavens, Jean-Luc Starck \& Alberto Krone-Martins, eds.}
\begin{document}

\maketitle

\begin{abstract}
The Large Sky Area Multi-Object Fiber Spectroscopic Telescope (LAMOST) project performed its five year formal survey since Sep. 2012, already fulfilled the pilot survey and the 1$^{st}$ two years general survey with an output - spectroscopic data archive containing about 3.5 million observations. One of the scientific objectives of the project is for better understanding the structure and evolution of the Milky Way. Thus, credible derivation of the physical properties of the stars plays a key role for the exploration. We developed and implemented the LAMOST stellar parameter pipeline (LASP) which can automatically determine the fundamental stellar atmospheric parameters (effective temperature $T_{\rm{eff}}$, surface gravity log \textit{g}, metallicity [Fe/H], radial velocity $V_{\rm{r}}$) for late A, FGK type stars observed during the survey. An overview of the LASP, including the strategy, the algorithm and the process is presented in this work.
\end{abstract}

\firstsection 
\section{Introduction}

\begin{figure}[b]
\begin{center}
 \includegraphics[width=\textwidth]{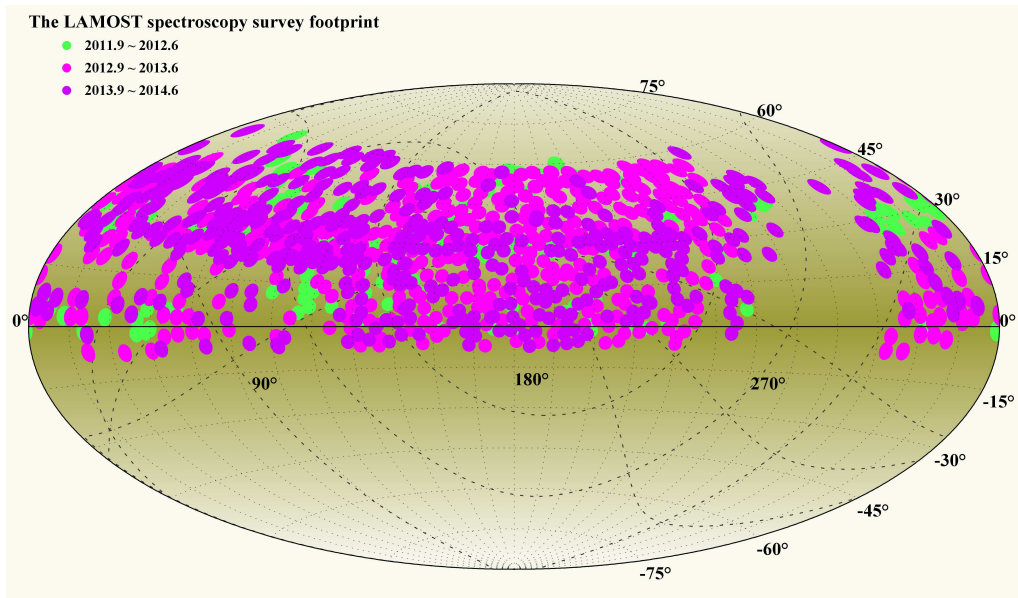}
 \caption{The LAMOST DR2 sky coverage.}
   \label{fig1}
\end{center}
\end{figure}

LAMOST is a unique reflecting Schmidt telescope, with a large aperture and a wide field of view (\cite[Cui et al. 2012]{Cui12}), sited at the Hebei province in China. It can simultaneously observe 4000 celestial objects and obtain the spectra through the multi-fiber system. After the previous two years commissioning survey and one year pilot survey, LAMOST accomplished its 1$^{st}$ two years general survey (\cite[Zhao et al. 2012]{Zhao12}) in the period between Sep. 2012 and Jun. 2014. The LAMOST First Data Release - DR1 (Pilot and the 1$^{st}$ year survey data) published in Aug. 2013, contains more than 2.2 million spectra, the majority of the targets are stars with a quantity of more than 1.9 million, others are Galaxy, QSO and UNKNOWN (low S/N). The whole DR2 (DR1 plus the 2$^{nd}$ year survey data, will be published at the end of 2014) sky coverage is demonstrated in the Fig.~\ref{fig1}. One of the two main components of the project is the LAMOST Experiment for Galactic Understanding and Exploration (LEGUE) which is focusing on the formation and evolution of the Milky Way (\cite[Deng et al. 2012]{Deng12}). The LAMOST data processing and analysis pipeline is integrated with three modulars: 2D pipeline (raw CCD data reduction, extraction, wavelength calibration), 1D pipeline (classification and the redshift estimation for galaxy and QSO, \cite[Luo et al. 2012]{Luo12}) and the LASP (\cite[Wu et al. 2011a]{Wu11a}).

\section{Data and Strategy}

The LAMOST spectra are with a resolving power of R$\sim$1800 covering 3800 - 9000 $\AA$ wavelength range. The consecutive data processing system of LAMOST is organized with a central versioned main data base (MDB) which is supported by MySQL. The observed data, the conventional observing information as well as all the intermediate and final output of each data reduction and analysis modular are all archived in this MDB, the published spectroscopic data are in the format of fits file. According to the moon status, we approximately divided the survey into three parts: dark night, bright night, and instrumental test night. The current selecting criteria for the LASP input are: `final class' is STAR and `final subclass' is late A or FGK type, with g band S$\slash$N $\geqslant$ 15 and S$\slash$N $\geqslant$ 6 for the bright and dark nights respectively.

\section{Methods and Procedure}

{\underline{\it CFI method}}. The name CFI is an abbreviation of `correlation function interpolation' (\cite[Du et al. 2012]{Du12}). Provided that the observed flux vector is \textit{O},  and the synthetic model flux vector is \textit{S}, theoretically, the best-fit pair is that cos$<$\textit{O},\textit{S}$>$ = 1. So it searchs for the best-fit by  maximizing the value of cos$<$\textit{O},\textit{S}$>$ as functions of $T_{\rm{eff}}$, log \textit{g} and [Fe/H], where cos$<$\textit{O},\textit{S}$>$ = (\textit{O}$\cdot$\textit{S})$\slash$($\mid$\textit{O}$\mid$$\times$$\mid$\textit{S}$\mid$) which is referred to as correlation coefficient. For the selection of the synthetic spectral girds, CFI employed Kurucz spectrum synthesis code based on the ATLAS9 stellar atmosphere model (\cite[Castelli \& Kurucz 2003]{CastelliKurucz03}). 

{\underline{\it ULySS method}}. ULySS (\cite[Koleva et al. 2009]{Koleva09}, \cite[Wu et al. 2011b]{Wu11b}) determines the atmospheric parameters via minimizing the $\chi^2$ between the observation and the model spectra which are generated by an interpolator built based on the ELODIE stellar library (Prugniel \& Soubiran 2001,2007). The model is:

$\rm Obs(\lambda) = P_{n}(\lambda) \times
G(\,\it{v},\sigma) \otimes \rm{TGM}(\,T_{\rm{eff}},log~$g$,\rm{[Fe/H]},\lambda)$, where $\rm Obs(\lambda)$ is the observed spectrum sampled in log$\lambda$, $\rm
P_{n}(\lambda)$ a series of Legendre polynomials of degree n, and
G(\,$\it{v}$,\,$\sigma$) a Gaussian broadening function parameterized by the residual velocity $v$, and the dispersion $\sigma$.
The multiplicative polynomial is meant to absorb errors in the flux calibration, Galactic extinction or any other source affecting the shape of the spectrum.

The TGM function is an interpolator of the ELODIE library, it consists of polynomial expansions of each wavelength element in powers of log($T_{\rm{eff}}$), log \textit{g}, [Fe/H] and f($\sigma$) (a function of the rotational broadening parameterized by $\sigma$). Three sets of polynomials are defined for three temperature ranges (roughly matching OBA, FGK, and M types) with important overlap between each other where they are linearly interpolated. 
The derived intrinsic external accuracy of ULySS for the FGK stars are 43\,K, 0.13\,dex, and 0.05\,dex for the individual parameters (\cite[Wu et al. 2011b]{Wu11b}). 

{\underline{\it Procedure}}. Based on the algorithms CFI and ULySS, by fitting the spectra, LASP executed in two stages to effectively derive the stellar parameters. Since the LAMOST spectral flux calibration is relative, in the 1$^{st}$ stage we measure the original observations, in the 2$^{nd}$ stage, we measure the normalized spectra. Considering the low instrumental response in both edges, as well as for optimizing the computing time and the storage space, CFI and ULySS respectively select [3850, 5500]\,$\AA$ and [4100, 5700]\,$\AA$ as the fitting window. In each stage, firstly, we utilize CFI to quickly get a set of initial coarse estimations, after that, by using CFI results as a guessed starting point, we adopt ULySS to obtain more accurate and credible measurements as the final output. CFI can rapidly select the appropriate stellar template fitting range, this advantage helps effectively reduce the computational quantity and time for ULySS by using less guessed starting grid.

\section{Conclusions}

The data procession phase of the large survey project comprises three important tasks: validation, calibration and in-mission software development. LASP is being improved and will continue producing an enormous set of parameters for various stars. Thus critical assessment of its measurements is significant for the realization of the project's scientific goal. In one of the independent external validations, by comparing the common hundreds of stars between LAMOST DR1 and the PASTEL Catalog (\cite[Soubiran et al. 2010]{Soubiran10}), we obtain precisions of 110\,K, 0.19\,dex, 0.11\,dex and 4.91\,km/s for $T_{\rm{eff}}$, log \textit{g}, [Fe/H] and $V_{\rm{r}}$ respectively in the specified temperature range (\cite[Gao et al. 2014]{Gao14}). A systemic comprehensive validation and calibration works according to different spectral types and SNRs have already been carried out, the results will be published in a separate work.

With the LAMOST DR1, a catalogue of stellar parameters which contains more than 1 million stars was simultaneous published, it has been characterized as the largest one in the world so far. For the spectra acquired during the 2$^{nd}$ year survey, LASP already preliminarily determined $\sim$\,0.8 million stars' parameters. Considering the five years operation, further LAMOST survey will yield more than 6 million stars covering in a wider sky area and depth, this will broaden the expected science especially for the characterization of the Galaxy with more various evolutionary stages of the stars collected.

{\underline{\it Acknowledgements}}. The authors thank the grants (Nos. 11103031, 11273026, 61273248, 11178021) from NSFC and the National Key Basic Research of China 2014CB845700.

\end{document}